\documentclass[a4paper,11pt]{article}
\usepackage[margin=1in]{geometry} 
\usepackage{times}
\usepackage{latexsym}

\usepackage{xcolor} 
\usepackage{colortbl}

\usepackage[utf8]{inputenc} 
\usepackage[T1]{fontenc}    

\usepackage[numbers,compress]{natbib}

\usepackage[colorlinks,citecolor=blue]{hyperref}         
\usepackage{url}            
\usepackage{booktabs}       
\usepackage{amsfonts}       
\usepackage{nicefrac}       
\usepackage{microtype}      

\usepackage{wrapfig}

\usepackage{graphicx}
\usepackage{todonotes}
\usepackage{amsmath}
\usepackage{amssymb}
\usepackage{amsthm}
\usepackage{array}
\usepackage{xcolor}
\usepackage{enumitem}
\usepackage{framed}
\usepackage{booktabs}
\usepackage{enumitem}
\usepackage{hyperref}
\usepackage{multirow}
\usepackage{multicol}
\usepackage{tabularx}

\usepackage[toc,page,header]{appendix}
\usepackage{minitoc}

\newcommand{\dataset}{\emph{SciEvo}\xspace}

\newcommand{\datasetbf}{\textbf{SciEvo}\xspace}

\usepackage{ifthen}
\usepackage{xspace}

\newboolean{arXivVersion}
\setboolean{arXivVersion}{true}  

\newcommand{\logo}{\raisebox{0ex}{\includegraphics[height=12pt]{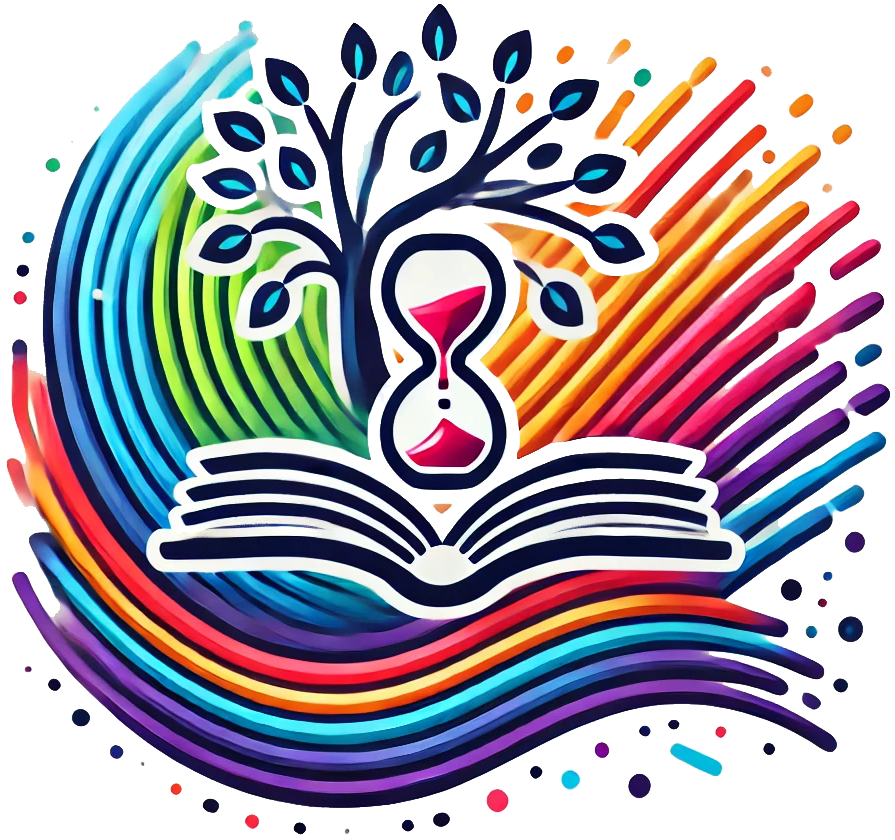}}\xspace}

\title{\dataset: A 2 Million, 30-Year Cross-disciplinary Dataset for Temporal Scientometric Analysis}

\author{
    Yiqiao Jin$^{1}$,
    Yijia Xiao$^{2}$,
    Yiyang Wang$^{1}$, 
    Jindong Wang$^{3}$ \\
    $^{1}$Georgia Institute of Technology, 
    $^{2}$University of California, Los Angeles, \\
    $^{3}$William \& Mary \\ 
    $^{1}$\texttt{\{yjin328,ywang3420\}@gatech.edu} \\
    $^{2}$\texttt{yijia.xiao@cs.ucla.edu} \\
    $^{3}$\texttt{jwang80@wm.edu}
}

\date{}

\begin{document}

\maketitle

\begin{abstract}
Understanding the creation, evolution, and dissemination of scientific knowledge is crucial for bridging diverse subject areas and addressing complex global challenges such as pandemics, climate change, and ethical AI. 
Scientometrics, the quantitative and qualitative study of scientific literature, provides valuable insights into these processes. 
We introduce \dataset, a longitudinal scientometric dataset with over two million academic publications, providing comprehensive contents information and citation graphs to support cross-disciplinary analyses. Using \dataset, we conduct a temporal study spanning over 30 years to explore key questions in scientometrics: the evolution of academic terminology, citation patterns, and interdisciplinary knowledge exchange. 
Our findings reveal critical insights, such as disparities in epistemic cultures, knowledge production modes, and citation practices. 
For example, rapidly developing, application-driven fields like LLMs exhibit significantly shorter citation age (2.48 years) compared to traditional theoretical disciplines like oral history (9.71 years). 
\ifthenelse{\boolean{arXivVersion}}{
}{
Our code and data are available at \url{https://anonymous.4open.science/r/Scito2M/}.
}
\end{abstract}

\vspace{-2mm}
\section{Introduction}
\label{section:intro}
\vspace{-1mm}

Scientific advances are crucial for 
addressing global challenges such as pandemics, energy security, climate change, social justice, and ethical AI. 
Tackling these issues requires a holistic understanding of how scientific knowledge evolves across disciplines. 
In this context, \textbf{scientometrics}--the qualitative and quantitative study of scientific literature--plays a pivotal role in uncovering the structure, dynamics, and evolution of research across fields~\citep{donthu2021conduct}. 
By analyzing publication contents and citation networks, scientometrics offers insights into key topics, trends, and scholars, providing valuable perspectives for researchers and policymakers in decision-making in scientific advances and solving global challenges. 


\noindent \textbf{Challenges.} Current scientometric studies face two major challenges: 
\emph{1) Limited Analytical Scope.} Despite extensive research in scientometrics--scrutinizing the creation~\citep{gu2021mapping}, diffusion~\citep{radev2012rediscovering}, and association~\citep{leto2024first} of academic knowledge--many studies focus on limited timespans~\citep{koch2021reduced,zhang2022investigating}, venues~\citep{ciotti2016homophily,jin2024agentreview}, or particular areas like natural language processing~\citep{radev2012rediscovering,singh2023forgotten,nguyen2024there} and human-computer interaction~\citep{oppenlaender2024past}. 
These gaps impedes a comprehensive understanding of critical issues, such as the \emph{breadth} (topical diversity) and \emph{depth} (long-term impact) of scientific knowledge exchange. 
\emph{2) Lack of Comprehensive Longitudinal Datasets.} 
While scientometric datasets are available, 
there is a scarcity of large-scale, longitudinal datasets that combine both content-level and citation-level information across multiple disciplines.

\noindent \textbf{This Work.} 
To address the lack of comprehensive datasets for such analyses, we present \datasetbf, a large-scale dataset of over 2 million academic papers 
from arXiv\footnote{https://arxiv.org/.} to support the study of \textbf{\underline{Sci}}entific \textbf{\underline{Evo}}lutions. In contrast to existing resources like S2ORC~\citep{lo-wang-2020-s2orc}, which require extensive preprocessing, filtering, and storage, \dataset is a \emph{ready-to-use} resource for comprehensive scientometric analyses, encompassing over 30 years since the inception of arXiv in 1991 and offers detailed metadata such as titles, abstracts, full-text\footnote{To comply with \href{https://info.arxiv.org/help/api/tou.html}{Term of Usage for arXiv}, we provide downloadable links instead of PDFs of paper e-prints. }, keywords, subject categories, and a comprehensive citation graph. 
\dataset also offers a suite of \textbf{analytic tools}, allowing users to perform detailed, longitudinal analysis of scientific knowledge evolution and citation patterns spanning multiple decades. Our content and citation analyses offer a holistic understanding of how interdisciplinary research contributes to solving global challenges. 
Our key findings are:
\begin{itemize}[leftmargin=1em]
\vspace{-1mm}
    \item \textbf{Paradigm Shifts}~\citep{shapere1964structure}. Scientific progress occurs through periodic \emph{leaps} rather than linear knowledge accumulation. Recent paradigm shifts have shown a noticeable change from theoretical to applied research.
    \item \textbf{Terminology Prominence.} Machine learning-related terms have seen a marked rise in prominence, accounting for an average of only 0.31 words of the top 20 annual terms prior to 2010, but surging to 9.5 words from 2015 onward.
    \item \textbf{Disciplinary Homophily}~\citep{zhang2018understanding}. Citation networks display a strong tendency towards homophily, with intra-disciplinary citation accounting for over 91\% of all citations. 
    \item \textbf{Epistemic Cultures}~\citep{cetina2007culture}. Different fields exhibit unique patterns in the production, validation, and citation of knowledge. Compared to applied research, basic research places greater emphasis on intra-disciplinary citations to maintain academic rigor and coherence.
    \item \textbf{Citation Amnesia}~\citep{singh2023forgotten}. Applied research exemplifies more citation amnesia and recency bias, favoring recent works while neglecting foundational, historical contributions. The median age of citation (AoC) for \texttt{LLM} research is $2.48$ years, compared with $9.71$ years for \texttt{Oral History}.
\end{itemize}

\noindent \textbf{Contributions.} Our contributions are three-fold:
\begin{itemize}[leftmargin=1em]
\setlength\itemsep{0em}
    \item \emph{Comprehensive Dataset.} We introduce \logo \dataset, a continuously updating dataset of over two million arXiv papers with detailed contents and citation data, 
    offering a valuable resource for longitudinal scientometric analysis across multiple disciplines. 
    \item \emph{Extensive Longitudinal Insights.} We perform a 30-year longitudinal analysis of academic literature,  
    offering new insights into how scientific knowledge is created and shared over time. 
    \item \emph{Analytic Tool Suite.} 
    We provide a set of analysis and visualization tools for scientometric research, allowing researchers to understand the  evolution of scientific terminology and citation. 
\end{itemize}

\begin{figure*}
    \centering
    \includegraphics[width=0.98\linewidth]{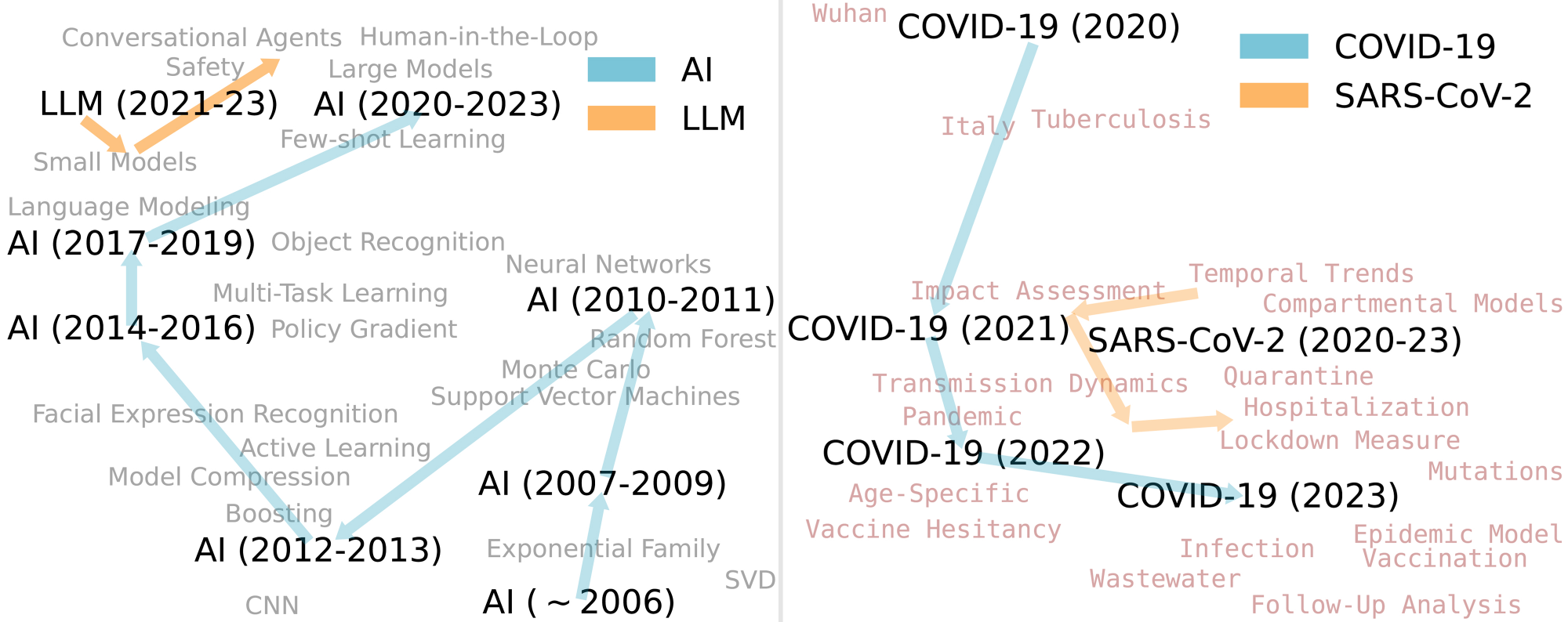}
    \vspace{-2mm}
    \caption{Keyword trajectories generated based on \dataset reflect critical paradigm shifts in AI and epidemiology research over time. Grey words represent t-SNE projections of keyword embeddings. (a) AI-related keywords; (b) COVID-related keywords. 
    }
    \label{fig:trajectory}
    \vspace{-4mm}
\end{figure*}


\section{The \dataset Dataset}
\label{sec:dataset}
\vspace{-2mm}

We select \href{https://arxiv.org/}{arXiv} as the data source since it has been a standard for disseminating preprints. Meanwhile, the permanence of arXiv papers ensures the integrity of the citation relations (more details in Appendix~\ref{app:data_selection}).

\noindent \textbf{Content Retrieval.} We retrieved all papers published on \href{https://arxiv.org/}{arXiv} from its establishment in 1991 to June 2024 that are under Creative Commons (CC) licenses using the \href{https://info.arxiv.org/help/api/index.html}{arXiv API}. 
The features of the papers include titles, abstracts, arXiv categories, comments, publishing and last updating timestamps, and full texts. 
We carefully curated the dataset to ensure broad representation across disciplines. Each paper is categorized into 8 subjects according to arXiv Category Taxonomy\footnote{\url{https://arxiv.org/category_taxonomy}}.

\noindent \textbf{Citation Retrieval.} As arXiv does not provide citation information, for each arXiv paper, we find the corresponding entry on semantic scholar, and retrieve the citation relations, publication venues, and author information using the semantic scholar API~\citep{kinney2023semantic}, which allow us to analyze the citation relations among papers from different subject areas.

\noindent \textbf{Keywords Extraction. }
Titles and abstracts in academic publications are typically crafted to highlight their most significant contributions, offering a concise yet accurate summary of the key concepts~\citep{krishnan2017unsupervised}. To enhance understanding of the paper contents, we extract keywords from each title and abstract using GPT-4o~\citep{openai2023gpt4}, inspired by previous works showing that LLMs demonstrate holistic understandings of academic literature~\citep{liang2024can}. 

\noindent \textbf{Statistics.} The resulting dataset contains 2.1 million papers spanning 34 years, falling under 8 groups and 156 categories. 
The dataset statistics is in Table~\ref{tab:stats} and the number of papers \& extract keywords per year is in Figure~\ref{fig:num_papers_and_keywords}. The detailed breakdown of the arXiv taxonomy is in Table~\ref{tab:arXivTaxonomy}.

\begin{table*}[!ht]
\centering
\setlength{\tabcolsep}{3pt}
\small
\begin{tabular}{llllllll}
\toprule
Dataset & Size & Text & Metadata & Citation & Tags & Analytic Tools & Disciplines \\
\midrule
Jurgens et al.~\citep{jurgens2018measuring} & 20,000 & $\checkmark$ & $\checkmark$ & $\checkmark$ & $\checkmark$ & $\checkmark$ & NLP \\ 
arXMLiv~\citep{ginev2020arxmliv} & 1.6M & $\checkmark$ & $\checkmark$ & ~ & ~ & ~ & multiple \\ 
SciXGen~\citep{chen2021scixgen} & 0.2M & $\checkmark$ & $\checkmark$ & ~ & ~ & ~ & CS \\ 
unarXive~\citep{saier2023unarxive} & 1.9M & $\checkmark$ & $\checkmark$ & $\checkmark$ & $\checkmark$ & ~ & multiple \\ 
arXiv Dataset~(\citeyear{arxiv_org_submitters_2024}) & 1.7M & $\checkmark$ & $\checkmark$ & ~ & ~ & & multiple \\ 
Ours & 2.1M & $\checkmark$ & $\checkmark$ & $\checkmark$ & $\checkmark$ & $\checkmark$ & multiple \\ 
\bottomrule
\end{tabular}
\caption{
The \dataset dataset offers the most comprehensive coverage across features, disciplines, and dataset sizes, surpassing other datasets in terms of breadth and depth.}
\label{tab:dataset_size}
\end{table*}




\vspace{-2mm}
\section{Diachronic Analysis of Terminology and Lexicons}
\vspace{-2mm}

\noindent \textbf{Thomas Kuhn's Theory of Paradigm Shifts} 
describes scientific progress as a series of periodic revolutions rather than a continuous, linear accumulation of knowledge~\citep{shapere1964structure}. Over time, existing research paradigms may become inadequate for address emerging problems, prompting the exploration of new, more effective approaches. 
To trace such shifts, we conduct diachronic analysis to study the evolution of language, concepts, 
and trends in academic literature over the years. 

\subsection{Macro-level Changes in Research Priority}
\label{sec:macro_level}

\begin{wrapfigure}{r}{0.6\textwidth}
\centering
\includegraphics[width=0.99\linewidth]{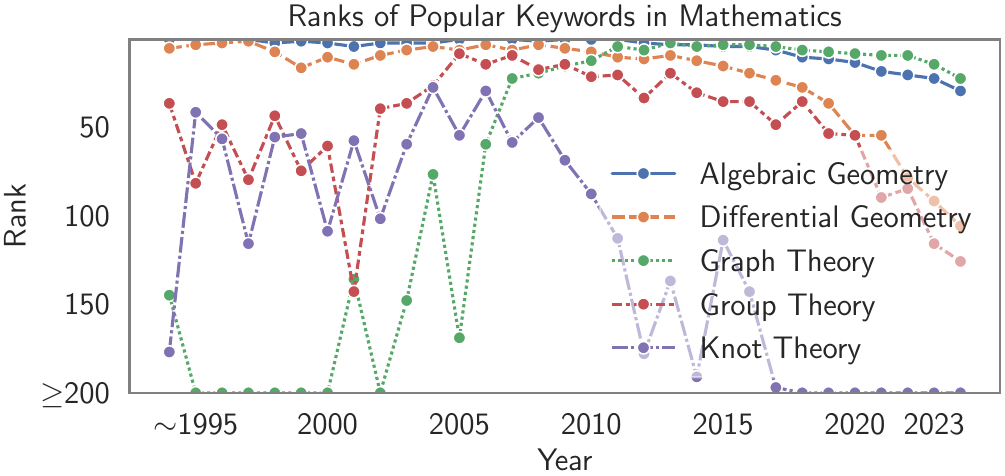}
\includegraphics[width=0.99\linewidth]{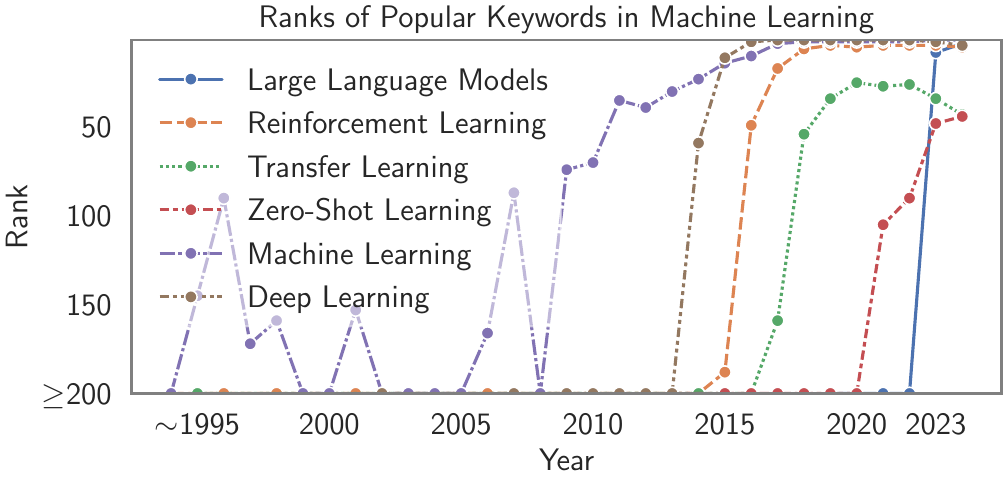}
\vspace{-2mm}
\caption{Evolution in the ranks of math and machine-learning terms among all keywords over time. Math keywords remain consistently popular but show a decline in the past decade, while ML keywords surged in prominence over the last ten years.
}
\label{fig:keyword_ranks_math_ml}
\vspace{-3mm}
\end{wrapfigure}

To trace paradigm shifts, we divide the papers into temporal snapshots according to their publication timestamps, and rank keywords in each snapshot by frequency. 
Figure~\ref{fig:keyword_ranks_math_ml} \& \ref{fig:keyword_ranks_all_others} illustrate how keyword prominence have changed shifts in algorithmic advances in recent years. 
From a macroscopic perspective, machine learning-related terms have significantly risen in prominence. Among the top 20 annual keywords, ML related keywords, accounting for an average of only 0.31 of the top 20 annual keywords prior to 10, but surging to an average of 9.5 from 2015 onward. 
In the 1990s, theoretical fields like \texttt{Quantum Field Theory} and \texttt{Particle Physics} were highly dominant and remain long-term popularity, 
as they were considered foundational research in advancing pure mathematics and theoretical physics. 
With the rise of computational technologies in the mid 2000s, data-driven methodologies such as Deep Learning and {Reinforcement Learning} constitute over 50\% of the top-10 keywords in the dataset, surpassing foundational keywords in mathematics like \texttt{Algebraic Geometry} and \texttt{Differential Geometry}. 
Keywords like \texttt{Deep Learning} and \texttt{Reinforcement Learning} showed notable growth from the 2010s onward, while more recent advancements such as \texttt{Large Language Models} and \texttt{Zero-shot Learning} gaining attention in the 2020s. 
The top-ranked keywords in each time period (Table~\ref{tab:most_mentioned_keywords}) confirms this shift and highlights the growing dominance of data-driven, AI-related research, which has overshadowed traditional theoretical fields in recent years. Keyword ranks for keywords in other subject areas such as chemistry, economics, and physics are in Appendix Figure~\ref{fig:keyword_ranks_all_others}.

\vspace{-2mm}
\subsection{From Foundations to AI and Societal Challenges}
\vspace{-2mm}
In computer science, the 1990s focus on foundational areas 
such as machine translation, computational complexity, and logic programming. 
These keywords indicate an emphasis on formal languages, the syntactic structure of language, and the mathematical foundations of computing. The presence of keywords like \texttt{TCP} and \texttt{ATM Networks} also highlight the prominence of network research. 
Starting in the early 2000s, keywords such as optimization, graph theory, and quantum computing emerged prominently, signifying a shift towards mathematical optimization. By the 2010s, deep learning gained prominence, with multiple emerging sub-topics such as neural networks, reinforcement learning, generative adversarial networks (GANs), and language models. This transformation aligns with the surge in interest in data-driven techniques, particularly the deep learning revolution that began reshaping fields such as computer vision, NLP, and robotics. Notably, in 2019 and beyond, terms like \texttt{Transformer}, \texttt{Self-supervised Learning}, and \texttt{Large Language Models (LLMs)} highlight the prominence of paradigms in generative AI, which have revolutionized NLP tasks and the broader AI landscape. 
New challenges and societal issues in 2020--2021 have driven recent research, with keywords such as \texttt{COVID-19}, \texttt{Blockchain}, and \texttt{Fairness} becoming more prominent. These terms reflect the increasing intersection of AI with pressing global issues like public health, data privacy, and decentralized technologies, indicating a shift towards more practical and ethical considerations in AI research. 

\vspace{-2mm}
\subsection{From Traditional Econometrics to Data-Driven Approaches}
\vspace{-2mm}
We examine keywords within specific subject areas based on the arXiv taxonomy (Table~\ref{tab:arXivTaxonomy}). 
Economics research from 2005 to 2024 shows a clear shift from traditional econometrics to data-driven, computational approaches. Early on (2005-2010), keywords like \texttt{Quantile Regression}, \texttt{Cornish-Fisher Expansion}, and \texttt{Panel Data} dominated, reflecting a focus on statistical modeling and policy evaluation. From 2010 onwards, machine learning techniques gained prominence, with terms like \texttt{Causal Inference}, \texttt{High-dimensional Models}, and \texttt{Nonparametric Methods}, signaling the increasing use of ML algorithms for economic analysis. 
By the mid-2010s, keywords such as \texttt{Game Theory}, \texttt{Social Networks}, and \texttt{Mechanism Design} reflected growing interest in strategic behavior and institutional design. By the 2020s, keywords like \texttt{COVID-19}, \texttt{Economic Growth}, and \texttt{Inequality} gained prominence, signaling a growing focus on societal issues and the integration of AI techniques in economic forecasting and decision-making. 

\vspace{-2mm}
\subsection{Micro-level Shifts in Term Usage}
\label{sec:micro_level}
\vspace{-3mm}

From a microscopic perspective, shifts in meanings and usage of academic keywords offer insights into how a field matures and how research priorities adapt to emerging challenges or technological advancements.
To trace such shifts, we analyze the co-occurrence patterns of terminology in title keywords of \dataset, as titles offer a high-level summary of paper content. 

\noindent \textbf{Embedding Training.} To capture \emph{temporality} in terminology usage, we partition the papers into temporal snapshots. For each snapshot, we construct a keyword co-occurrence graph using the extracted keywords in \dataset (Section~\ref{sec:dataset}). Each paper serves as a hyperedge connecting all keywords associated with its title. 
On average, each snapshot includes 9,028 keywords. 
We train a two-layer Graph Convolutional Network (GCN)~\citep{kipf2022semi} model with a link prediction objective on each snapshot to extract the high-order co-occurrence relations among keywords. To ensure quality of the embeddings, we filter the embeddings and keep phrases that appear $\ge 3$ times, following\ifthenelse{\boolean{arXivVersion}}
{~\citep{hamilton2016diachronic}}
{~\citep{hamilton2016diachronic}}.

\noindent \textbf{Temporal Embedding Alignment} 
To compare word vectors across different time periods, we align them to the same embedding space using orthogonal Procrustes~\citep{ten1977orthogonal}, which effectively preserve proximity of relevant terms~\citep{hamilton2016diachronic}. 
Let $\mathbf{E}_{t} \in \mathbb{R}^{d \times|V|}$ be the word embedding matrix at year $t$, where $d$ is the embedding dimension and $V$ is the vocabulary, we align these embeddings by optimizing: 
\begin{equation}
    \mathbf{R}_{t}= \underset{\mathbf{Q}^{\top} \mathbf{Q}=\mathbf{I}}{\operatorname{argmin}}\left\|\mathbf{E}_{t} \mathbf{Q} - \mathbf{E}_{t+1}\right\|_F,
\end{equation} 
where $\mathbf{R}_{t}$ is the rotation matrix for orthogonal transformation that best aligns $\mathbf{E}_{t}$ to $\mathbf{E}_{t+1}$. $\mathbf{Q}$ is an orthogonal matrix that preserves the geometric properties of the embeddings. 
A \emph{keyword trajectory} traces the movement of a keyword in the embedding space over time~\citep{jin2024empowering}. The trajectory of a keyword $w$ converge to a word $w'$ in the embedding space if $w$ and $w'$ frequently appear in similar contexts over a given time period. 
To interpret the trajectories, we project the keyword embeddings and trajectories into 2D space using t-SNE~\citep{van2008visualizing}. 

The keyword trajectory of \texttt{Artificial Intelligence} in Figure~\ref{fig:trajectory}a illustrates important paradigm shifts within AI research over the past few decades.  
The research focus transitions from early 
theoretical foundations such as \texttt{Exponential Family} and \texttt{SVD} to core ML techniques like \texttt{Support Vector Machines} and \texttt{Monte Carlo} (2007 -- 2009).  
With growing computational power and larger datasets, more complex models like \texttt{CNN} and \texttt{Boosting} (2012–2013) emerged, followed by advanced tasks like \texttt{Multi-Task Learning} and \texttt{Active Learning} (2014–2016).
As the field mature, attention turned to applications such as \texttt{Object Recognition} in 2017 and \texttt{Few-shot Learning} around 2020. 
Recent movements towards \texttt{Large Models} and \texttt{Human-in-the-Loop} reflect new challenges in scalability and human intervention.

\begin{figure*}[t]
\centering
\vspace{-2mm}
\includegraphics[width=0.97\linewidth]{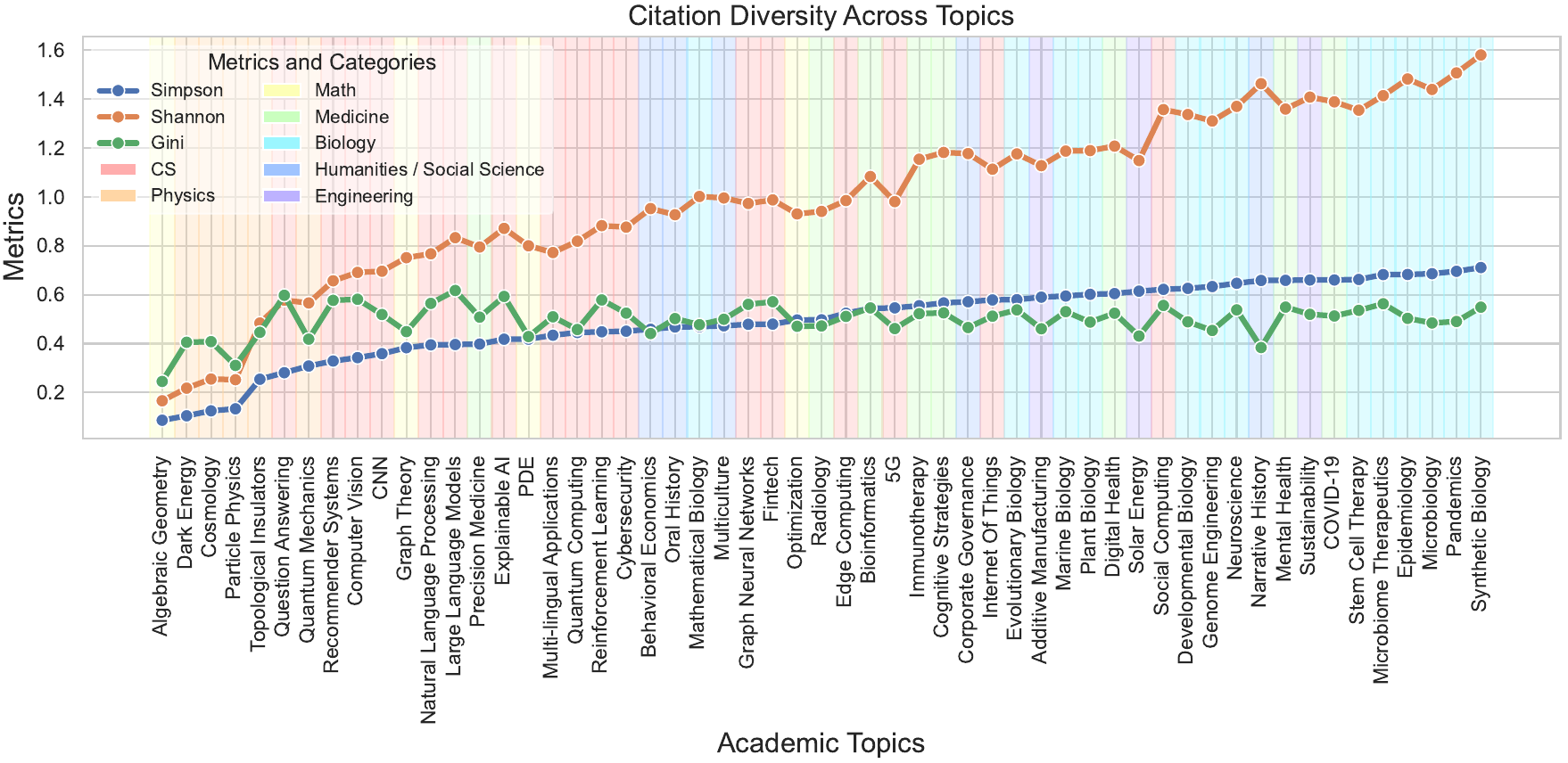}
\caption{Citation diversity in terms of Simpson's Diversity Index, Shannon Diversity Index, and Gini Index. Topics are sorted according to Simpson's Diversity Index. The background is colored according to subject areas. Higher values for $\mathrm{Simpson}(i), \mathrm{Shannon}(i)$ and lower values for $\mathrm{Gini}(i)$ imply greater diversity. 
}
\label{fig:citation_diversity}
\vspace{-5mm}
\end{figure*}

\noindent \textbf{Co-Evolution of Technology and Society.} Technological advancements and societal changes mutually influence each other. 
The overlapping trajectories between \texttt{AI} and \texttt{LLM} show how AI research has increasingly incorporated large-scale models to address societal needs. 
The trajectory for \texttt{Large Language Models} (\texttt{LLM}), after its first occurrence around 2021, starts with technical topics like \emph{language modeling} and is sometimes discussed in tandem with \texttt{Small Models}. 
As LLMs gained prominence, their associated terms evolved to reflect their broader academic adoption, incorporating keywords like \emph{Human-in-the-Loop}, \emph{Safety}, and \emph{Large Models}, reflecting increased societal concerns and the need for responsible, interpretable, and user-centric AI development with human oversight as LLMs gained prominence. 
Notably, \texttt{LLM} represents a recent topic, thus remaining in a confined region of the embedding space filled with technical terminologies. while \texttt{AI} spans a wider scope, covering decades of research.

For each pair of keywords, the cosine similarity between their embeddings serves as a proxy for their lexical association~\citep{pecina2010lexical} in academic discourse. 
Table~\ref{tab:closest_ml_keywords} presents the phrases with the highest cosine similarity with \texttt{Machine Learning} across different time periods, highlighting the evolution of ML research focus from \texttt{Neural Networks} in the mid-1990s, \texttt{Machine Translation} in the early 2000s, followed by \texttt{Reinforcement Learning} in the 2010s, \texttt{Ethics} and \texttt{Scalability} concerns in the 2020s, and in recent years, \texttt{Large Language Models (LLMs)} and \texttt{Conversational Agents}. 
This progression reflects how the focus of machine learning research has evolved in response to societal needs. 

\begin{wrapfigure}{r}{0.42\textwidth}
\centering
\includegraphics[width=0.97\linewidth]{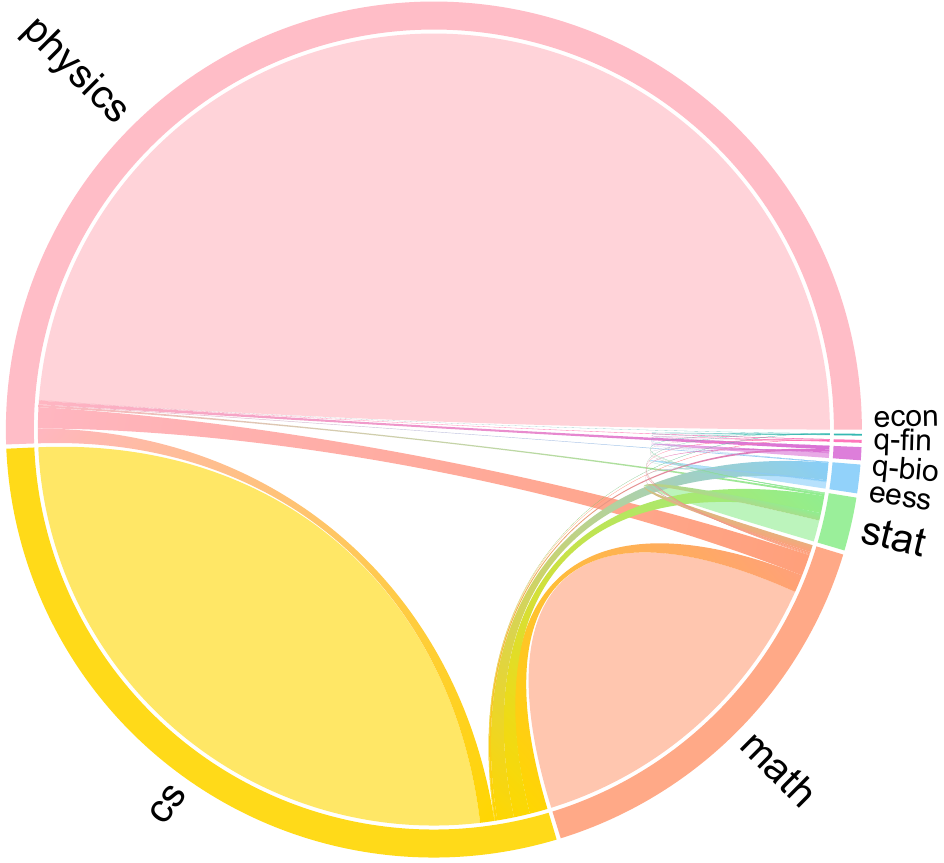}
\vspace{-2mm}
\caption{Literature in \dataset exhibits higher intra-disciplinary than cross-disciplinary citations.}
\label{fig:cross_disciplinary_citation}
\vspace{-2mm}
\end{wrapfigure}

\noindent \textbf{Near Synonymy and Lexical Choice.} 
Near-synonymy refers to the relationship between words with similar meanings~\citep{edmonds2002near}.
For example, \texttt{SARS-CoV-2} refers to the virus or pathogen that causes the disease known as \texttt{COVID-19}. 
These terms are sometimes used interchangeably as \textbf{denominative variants} that refer to the same concept in non-technical contexts~\citep{benitez2023denominative}. 
Trajectories of the two keywords 
in Figure~\ref{fig:trajectory}b demonstrate their lexical choices across different contexts. 
As a technical term, \texttt{SARS-CoV-2} is primarily used by biomedical professionals in academic contexts to ensure precision in scientific communication. 
The keyword follows a constrained trajectory, primarily associated with scientific discussions in epidemiology, virology, and healthcare as it intersect with keywords such as \texttt{Compartmental Models}, \texttt{Temporal Trends}, and \texttt{Hospitalization}. 
In contrast, \texttt{COVID-19} is a more 
widely recognized and accessible term that extends beyond medical contexts for 
clear communication across diverse fields such as politics, sociology, and economics. Its trajectory reflects broader societal concerns, 
with early associations including geographical keywords (\texttt{Wuhan}, \texttt{Italy}, and \texttt{India}) reflecting a continued efforts in tracking the outbreak. Later years (2021-22) shift towards public health measures like \texttt{Lockdowns} and \texttt{Quarantine}, reflecting the socio-political impacts of the pandemic. 

\begin{figure*}[t]
\centering
\includegraphics[width=0.98\linewidth]{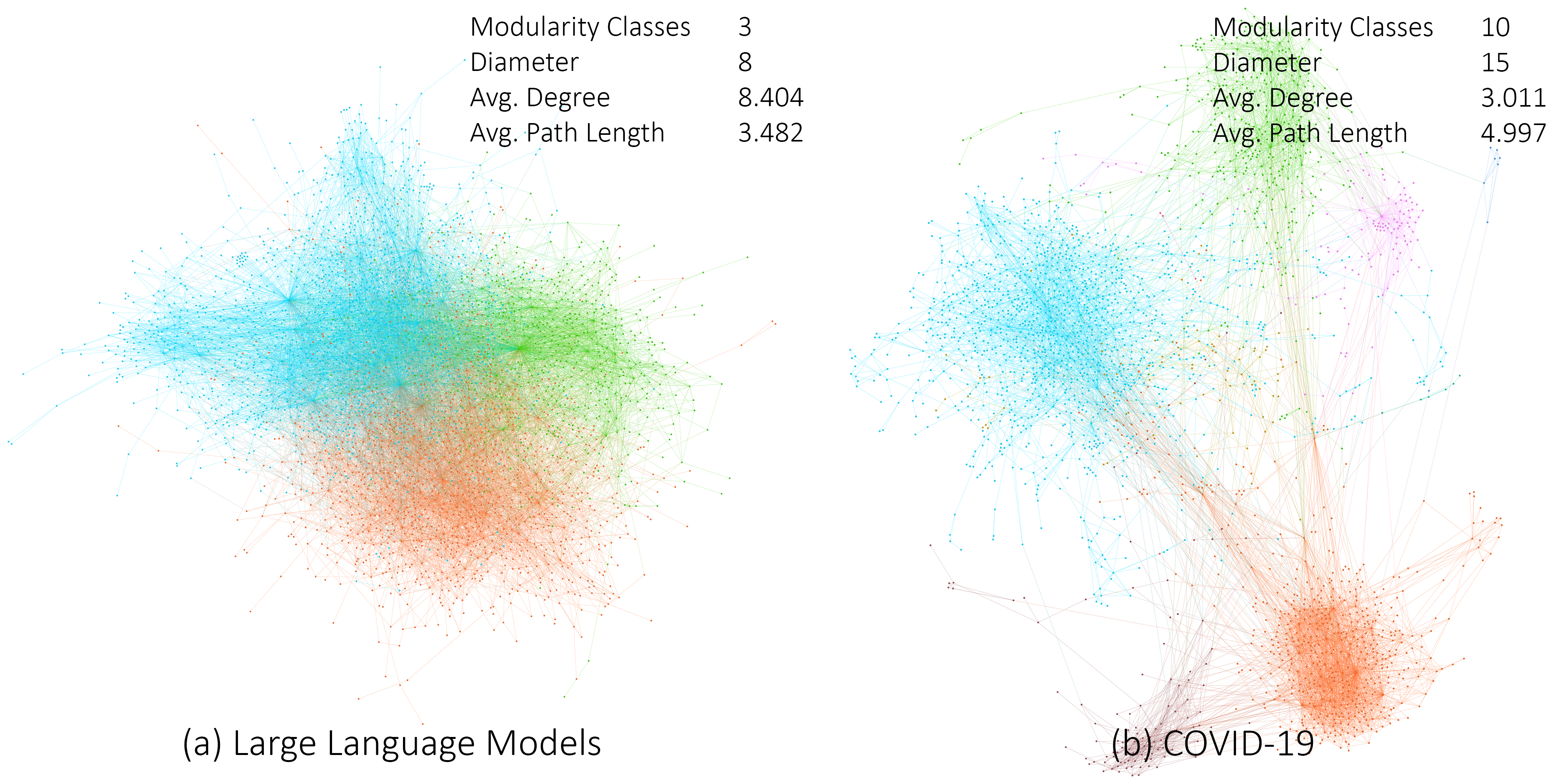}
\vspace{-2mm}
\caption{Citation graphs of LLM (left) and epidemiology (right) literature show distinct citation patterns. LLM publications is densely connected with three modularity classes forming a large cluster, suggesting a highly specialized, self-referential field. The epidemiology graph, with 10 modularity classes, highlights several distinct research subfields that reflect the interdisciplinary nature of the field.}
\label{fig:citation_graph}
\vspace{-3mm}
\end{figure*}

\vspace{-2mm}
\section{Topical \& Temporal Citation Dynamics}
\vspace{-2mm}


Citations form the backbone of scientific inquiry by connecting current research with prior contributions~\citep{boyack2005mapping}. The diversity of citations is essential as it fosters comprehensive understandings of research problems and encourages interdisciplinary collaboration. In this section, we analyze two key aspects of citations: topical diversity, which reflects the \emph{breadth} of domains a paper engages with, 
and temporal diversity, which reveals the \emph{depth} of references across time periods. 

\vspace{-2mm}
\subsection{Topical Diversity}
\label{sec:breadth}
\vspace{-1mm}

The topical diversity reflects the \emph{breadth} of citation. 
A higher topical diversity suggests that the paper draws on insights from a broader range of disciplines, fostering cross-disciplinary innovation and perspectives. To measure this, we use three well-established metrics: Simpson's Diversity Index ($\mathrm{Simpson}(i)$), Shannon's Diversity Index ($\mathrm{Shannon}(i)$), and Gini Index ($\mathrm{Gini}(i)$). Details can be found in Appendix~\ref{app:metrics_topical_diversity}. 
\begin{figure*}[h]
    \centering
    \includegraphics[width=0.97\linewidth]{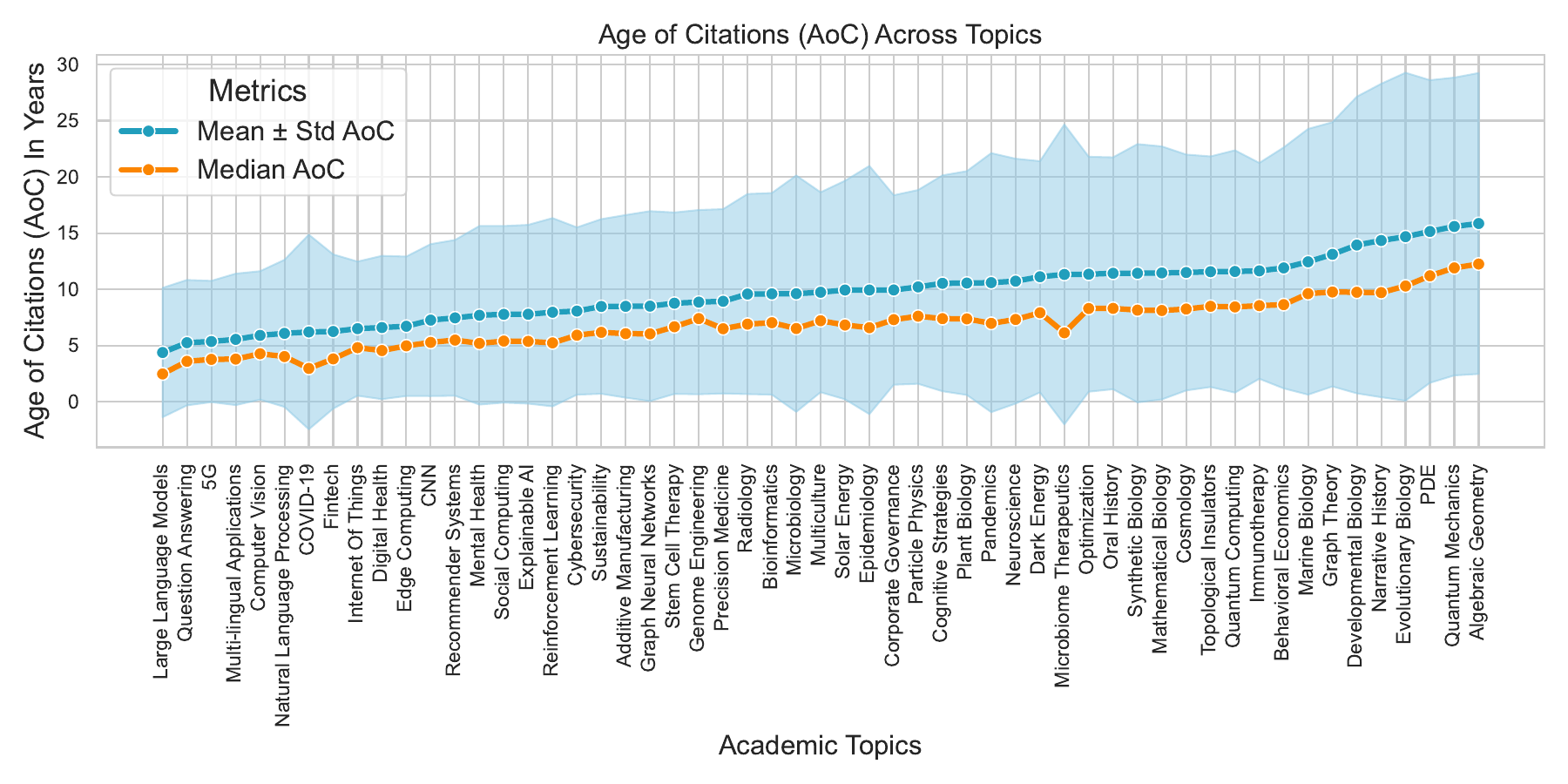}
    \vspace{-3mm}
    \caption{Age of Citation (AoC) of papers under each academic topic.}
    \label{fig:AoC_by_Topics}
    \vspace{-3mm}
\end{figure*}

\noindent \textbf{Homophily}~\citep{zhang2018understanding}. Networks of academic knowledge sharing often exhibit \emph{homophily}, where papers preferentially connect with works from their own discipline~\citep{csimcsek2008navigating,ciotti2016homophily}. 
The intra- and cross-disciplinary citation in Figure~\ref{fig:cross_disciplinary_citation} show that over 91.0\% citations occur within the same discipline. 
This \emph{homophily} stems from researchers' tendency towards shared methods, expertise, language, and conceptual frameworks within an academic field. 


\noindent \textbf{Epistemic Culture} refer to the distinct ways in which knowledge is produced, validated, and shared within different scientific disciplines~\citep{cetina2007culture}. 
According to 
\ifthenelse{\boolean{arXivVersion}}
{~\citep{cetina2007culture}}
{~\citep{cetina2007culture}}, 
global scientific knowledge production is characterized by \emph{disunity}, where each subject area operates under its distinct epistemic culture. 
To examine these differences, 
We calculate citation diversity across academic topics (Figure~\ref{fig:citation_diversity}), revealing how epistemic cultures shape the interdisciplinary scope and maturity of research fields. 
More theoretical or \emph{pure} disciplines, such as \texttt{Algebraic Geometry} in mathematics and \texttt{Quantum Mechanics}, \texttt{Dark Energy}, and \texttt{Particle Physics} in physics, exhibit an epistemic culture with an internal focus. 
These fields rely on well-established, codified knowledge practices that prioritize internal citations, reinforcing disciplinary coherence while limiting interdisciplinary input from external fields. 
Conversely, emerging areas like \texttt{Large Language Models} (LLMs) and \texttt{Digital Health} demonstrate broader interdisciplinary citation patterns. As these areas are still developing their intellectual foundations, they draw on a wide range of research to shape their evolving epistemic frameworks. 

\begin{wrapfigure}{r}{0.56\textwidth}
\centering
\includegraphics[width=0.97\linewidth]{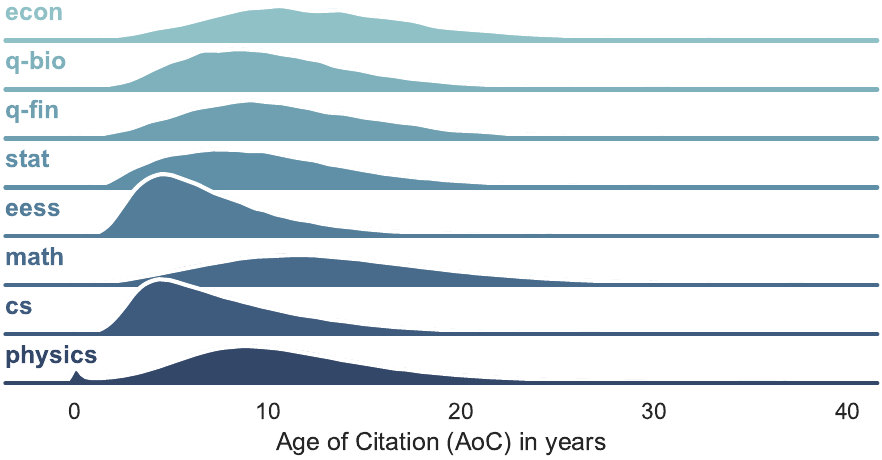}
\caption{
Age of Citation (AoC) across various \href{https://arxiv.org/category_taxonomy}{arXiv subjects} shows distinct trends. \texttt{eess} and \texttt{cs} 
exhibit left-skewed distributions, indicating a preference towards recent citation. In contrast, disciplines such as \texttt{physics}, \texttt{math}, and \texttt{econ} demonstrate broader AoCs, reflecting their reliance on historical foundational research.
}
\label{fig:aoc_all_subjects}
\vspace{-1mm}
\end{wrapfigure}

\noindent \textbf{Knowledge Production Modes and Field Theory.} 
\ifthenelse{\boolean{arXivVersion}}
{\citep{gibbons1994new}}
{\citep{gibbons1994new}}
conceptualized two modes of knowledge production. Mode 1 refers to basic research driven by fundamental principles and theories, as seen in physics and math. 
Mode 2 is problem-oriented, usually associated with applied research that requires interdisciplinary collaboration, 
as exemplified by \texttt{Graph Neural Networks} (GNNs) and \texttt{Quantum Computing}. These fields align with Mode 2, involving higher contributions from prior research ($\mathrm{Simpson}(i) = 0.479 / 0.445$) compared to theory-centric topics like \texttt{Graph Theory} and \texttt{Quantum Mechanism} ($\mathrm{Simpson}(i) = 0.383 / 0.308$). 
From the perspective of \textbf{field theory}~\citep{bourdieu1983field}, academic disciplines operate like competitive \emph{social fields}, where scholars vie for intellectual capital. 
Established fields like \texttt{Graph Theory} have a highly structured system of knowledge production with a stable, widely accepted body of references. Scholars are expected to operate within a tightly regulated intellectual space, where contributions are expected to align closely with established theories.
In contrast, emerging fields like
\texttt{Graph Neural Networks} operate in less structured intellectual spaces, with scholars drawing upon a diverse array of disciplines, such as computer science, machine learning, and domain-specific fields like social network analysis and recommender systems~\citep{li2022meta,li2023mhrr}. 
Similarly, interdisciplinary topics such as \texttt{Social Computing} and \texttt{COVID-19} exhibit greater diversity across metrics as they integrate insights from multiple knowledge domains--sociology, epidemiology, 
computer science, and network science--to address complex, multi-faceted problems. 

\begin{figure*}[htbp]
\centering
\includegraphics[width=0.98\linewidth]{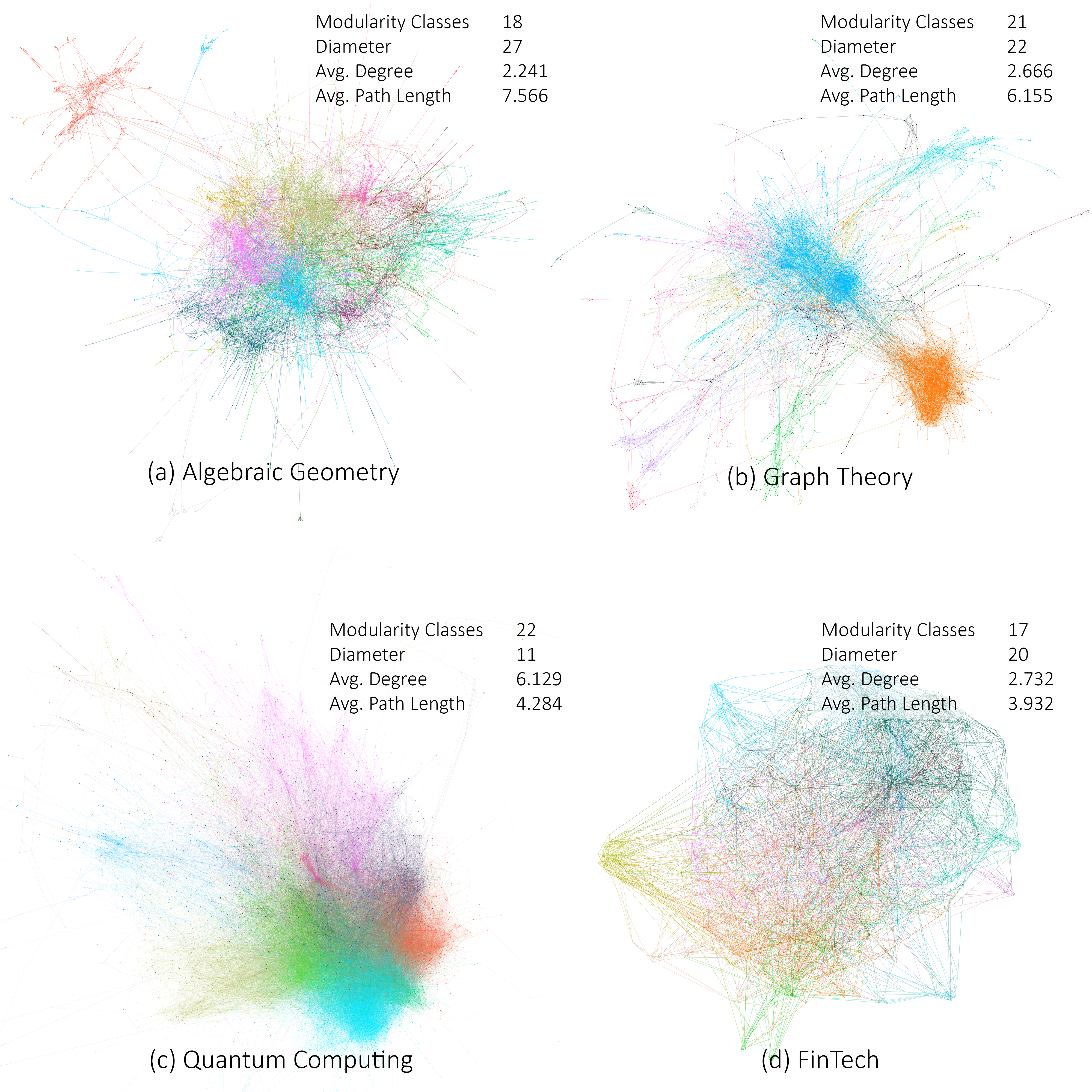}
\vspace{-5mm}
\caption{Citation graphs of a) Algebraic Geometry, b) Graph Theory, c) Quantum Computing, and d) FinTech. Graph Theory shows two relatively dominant, independent areas of focus, while Quantum Computing exhibits strong interconnections with diverse research communities.}
\label{fig:citation_graph_other_topics}
\vspace{-5mm}
\end{figure*}

\vspace{-3mm}
\subsection{Temporal Diversity}
\label{sec:depth}
\vspace{-2mm}
Temporal diversity measures the distribution of citations across different time periods, reflecting the \emph{depth} and profound influence of academic works over time. 
Examining temporal diversity can uncover issues like \emph{citation amnesia}~\citep{singh2023forgotten} and \textbf{Recency Bias}~\citep{abah2016recency}, the tendency to prioritize recent contributions and overlook significant historical knowledge.

\noindent \textbf{Pace of Innovation} refers to the speed of technological advancement, which varies significantly across academic fields. We quantify this pace using the Age of Citation (AoC)~\citep{singh2023forgotten}, defined as the publication time gap between a paper and its cited works. 
\texttt{eess} (Electrical Engineering and Systems Science) and \texttt{cs} (Computer Science) exhibit right-skewed AoC distributions (Figure~\ref{fig:aoc_all_subjects}), indicating 
a rapid pace of development driven by recent research. 
This is especially evident in fast-evolving topics like \texttt{Natural Language Processing} and \texttt{Machine Learning} (Figure~\ref{fig:AoC_by_Topics}), which show median AoCs of $4.02$ and $5.06$ years. In contrast, humanities subjects like \texttt{Narrative History} and \texttt{Oral History} have significantly higher median AoCs of $8.31$ and $9.71$ years. 
Disciplines such as \texttt{econ} and \texttt{math} display relatively flat AoC distributions, suggesting a holistic development process that relies on both recent innovations and long-established foundational works. 

\vspace{-0mm}
\noindent \textbf{Citation Network Structures. } 
Figure~\ref{fig:citation_graph} \& \ref{fig:citation_graph_other_topics} display citation graphs across various research topics, where nodes represent papers, edges represent citation relations, and colors indicate communities identified via the Louvain Community Detection Algorithm~\citep{blondel2008fast}. 
The LLM literature (Figure~\ref{fig:citation_graph}a) displays a densely interconnected network with three major modularity classes forming a dominant cluster. This suggests a rapid and concentrated \textbf{diffusion of innovation}~\citep{rogers2014diffusion} as researchers converge on core methodologies, creating a relatively unified body of knowledge. 
The rapid innovation may be driven by strong technological momentum and the lower barriers to entry, though concerns about the oversight of social impacts remain. 
The epidemiology literature (Figure~\ref{fig:citation_graph}b) exhibits a fragmented structure with weak ties linking distinct cluster. Each cluster represents research on specific outbreaks or frameworks, often spanning multiple decades. 
While allowing for domain-specific innovations, it complicates the synthesis of knowledge across the broader field.

\vspace{-2mm}
\section{Related Works}
\vspace{-2mm}

\noindent \textbf{Scientometrics} plays a crucial role in understanding the structure and evolution of scientific research, 
providing insights into how knowledge is produced, disseminated, and consumed across academic communities~\citep{wahle2023we,wahle2024citation}. 
Previous scientometric studies explored aspects such as interdisciplinary knowledge associations~\citep{leto2024first,thilakaratne2018automatic}, academic knowledge diffusion~\citep{jin2023predicting,jin2024empowering}, concept extraction~\citep{krishnan2017unsupervised}, academic community factions~\citep{sim2012discovering}, figures~\citep{li2024multimodal}, research artifacts usage patterns~\citep{koch2021reduced}, and estimation measures of research impact~\citep{radev2012rediscovering}. 

\noindent \textbf{Citation analysis}, in particular, is widely used to measure impact of researchers, publications, institutions, and venues across disciplines. 
While studies have investigated dimensions such as citation polarity~\citep{ghosh2018determining,radev2012rediscovering}, purpose~\citep{jha2017nlp}, and influence~\citep{gao2013spatiotemporal}, most existing works fail short in cross-disciplinary scope~\citep{bollmann2020forgetting,bird2008acl} 
and temporal depth~\citep{koch2021reduced,zhang2022investigating}. 

\vspace{-.1in}
\section{Conclusion}
\vspace{-.1in}
We proposed \dataset, a 2 million, 30-year dataset for longitudinal scientometric analysis. 
By leveraging the data, researchers can explore shifts in epistemic priorities, paradigm changes, and citation amnesia across multiple fields. 
Its extensive coverage and analytic tools empower researchers to better understand the creation, dissemination, and application of scientific knowledge.

\clearpage

\bibliographystyle{unsrt}
\bibliography{cite}

\appendix


\section{Limitations}
\label{app:limitations}
We acknowledge the following limitations about \dataset. 
First, while arXiv is widely adopted and represents a significant portion of academic literature, it may not fully reflect research published in local venues or non-English conferences that are less closely associated with arXiv, potentially leading to gaps in coverage. 
Second, citation retrieval relies on Semantic Scholar data, which, while generally robust, may occasionally lack metadata for very recent papers or those in specialized areas. Lastly, the arXiv taxonomy can sometimes group diverse subfields, complicating the task of distinguishing closely related research areas. 
Future enhancements can integrate additional data sources such as CrossRef\footnote{\url{https://www.crossref.org/}} and Google Scholar\footnote{\url{https://scholar.google.com/}} to enhance topical \& citation coverage and subject classification. 
Expanding the temporal scope to include real-time updates would also ensure the dataset's ongoing relevance, supporting longitudinal studies on shifting research trends, especially in response to major global events or scientific breakthroughs. Future works can also expand the coverage of papers into broader subject areas, such as humanities.

\section{Ethical Considerations}
\label{sec:ethics}
\noindent \textbf{Compliance with Data Usage Policies.} 
We are committed to ensuring that all data collection and analyses strictly comply with Terms of Use for arXiv API\footnote{\url{https://info.arxiv.org/help/api/tou.html}}, Semantic Scholar API\footnote{\url{https://www.semanticscholar.org/product/api/license}}, and Ai2 Privacy Policy\footnote{\url{https://allenai.org/privacy-policy/2022-07-21}}. 
\begin{wraptable}{r}{0.5\textwidth}
\centering
\begin{tabular}{l|c}
\toprule
\textbf{Feature} & \textbf{Statistics} \\
\midrule
\#Papers & 2,118,385 \\
Time Span & 1991 -- 2024 \\
\#Groups & 8 \\
\#Categories & 156 \\
Avg. \#Categories & 1.98 $\pm$ 1.05 \\
Avg. Length (Title) & 10.58 $\pm$ 4.07 \\
Avg. Length (Abstract) & 146.04 $\pm$ 53.70 \\
Avg. \#Keywords (Title) & 3.06 $\pm$ 0.30 \\
Avg. \#Keywords (Abstract) & 14.62 $\pm$ 1.41 \\
\bottomrule
\end{tabular}
\caption{Statistics of the dataset. \texttt{Avg. Length (Title/Abstract)} are the average numbers of words in the titles \& abstracts, respectively. \texttt{Avg. \#Keywords (Title/Abstract)} are the average numbers of LLM-extracted keywords from each paper's title and abstract. }
\label{tab:stats}
\vspace{-1mm}
\end{wraptable} 
The dataset is used strictly for academic research purposes, focusing on scientometric analysis, and adhering to all data privacy and intellectual property guidelines. 
The dataset usage aligns with recommended academic use cases, focusing on the retrieval and analysis of scientometric data for research. We strictly follow ethical guidelines concerning access, data privacy, and intellectual property. Importantly, \dataset should not be used for unfair or harmful evaluations of individual researchers, especially those from underrepresented groups or early-career scholars. Ethical usage should prioritize promoting open science, transparency, fairness, and responsibility within the academic community.

\noindent \textbf{Path Dependency.}~\citep{page2006path} 
Our results show that academic research is often shaped by historical trajectories, where early dominant paradigms influence future directions (a concept known as \emph{path dependency})~\citep{page2006path}. This can result in a \emph{lock-in} effect, where unconventional ideas or new directions that diverge from established patterns are less likely to gain traction. 
This trend is reinforced by institutional practices, funding agencies, and high-impact journals, which tend to favor research that builds on well-established theories. Future investigations can explore how these historical and institutional factors limit the diversification of research.  
\noindent \textbf{Citation Patterns and Their Implications.} Our findings in citation diversity have profound implications for research development and interdisciplinary collaboration. 

Citation \textbf{breadth}--the range of topics referenced--serves as an indicator of interdisciplinarity. A higher citation breadth suggests that a field integrates diverse knowledge, which is crucial for addressing complex global challenges. Conversely, a low citation breadth indicates a focus on a narrow set of foundational theories or methods. While this specialization can lead to deep expertise, it may also risk intellectual isolation and hinder innovation. 

In terms of citation \textbf{depth} -- the temporal span of cited works -- fields like natural language processing (NLP) demonstrate \emph{citation amnesia}. Approximately 62\% of citations point to works published within the last five years, and only about 17\% are works older than ten years~\citep{singh2023forgotten}. 
Our research has highlighted the prevalence of this problem and underscored the need to strike a balance between maintaining relevance with recent work and preserving a strong connection to foundational theories.


\noindent \textbf{Cognitive Load and Information Availability.}
The growing volume of publications poses a significant cognitive load on researchers, making it challenging to stay current with all relevant literature. To cope with this overload, researchers may rely on cognitive shortcuts, such as the \textbf{availability heuristic}~\citep{schwarz1991ease}, favoring recent or highly visible information over older, less prominent work. Future research should consider the impact of these cognitive biases on citation practices and the potential narrowing of scholarly discourse.


\section{Experimental Details}
\label{app:experimental_details}

\subsection{Metrics for Topical Diversity}
\label{app:metrics_topical_diversity}
\noindent \textbf{Metrics.} 
To quantify topical diversity, we measure the proportion of references associated with each academic field. Let $R(i)$ denote the set of references for a paper $i$. Each reference $j$ is associated with one or more subject areas denoted by $F(i, j)$. Here, we use the Field of Study attribute in semantic scholar~\citep{kinney2023semantic}. If a reference belongs to multiple fields, each subject is credited with an equal fraction of the reference's contribution. The total contribution of a subject $s$ to paper $i$ is: 
\begin{equation}
    C(i, s) = \sum_{j=1}^{|R(i)|}  \frac{\delta_s(j)}{|F(i, j)|}
\end{equation}
where $\delta_s(j)$ is an indicator function that equals $1$ if subject $s$ is in $F(i, j)$, and $0$ otherwise. 
Topical diversity is then measured using three well-established indices: Simpson's Diversity Index ($\mathrm{Simpson}(i)$), Shannon's Diversity Index ($\mathrm{Shannon}(i)$), and Gini Index ($\mathrm{Gini}(i)$): 
\begin{equation}
    \mathrm{Simpson}(i) = 1 - \sum_{j=1}^{k_i} C(i, j)^2,
\end{equation}
\begin{equation}
    \mathrm{Shannon}(i) = -\sum_{j=1}^{k_i} C(i, j) \log C(i, j),
\end{equation}
\begin{equation}
    \mathrm{Gini}(i) = \frac{\sum_{s=1}^{k_i} \sum_{t=1}^{k_i} |C(i, s) - C(i, t)|}{2 k_i \sum_{s=1}^{k_i} C(i, s)}.
\end{equation}

$\mathrm{Simpson}(i)$ measures the probability that two randomly selected references belong to different subjects. $\mathrm{Shannon}(i)$ quantifies the uncertainty in predicting the subject area of a randomly selected reference. 
$\mathrm{Gini}(i)$ assesses the inequality in the distribution of references across subject areas.

\begin{figure*}[htbp]
    \centering
    \includegraphics[width=0.49\linewidth]{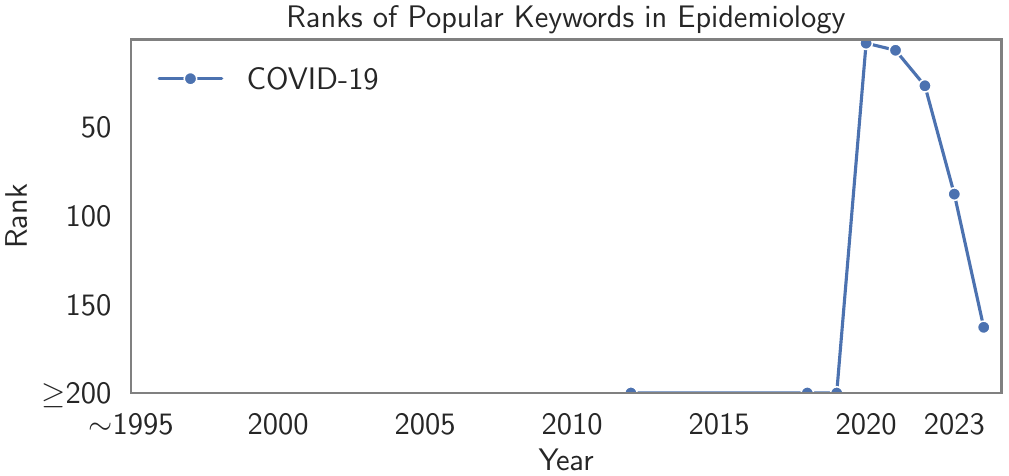}
    \includegraphics[width=0.49\linewidth]{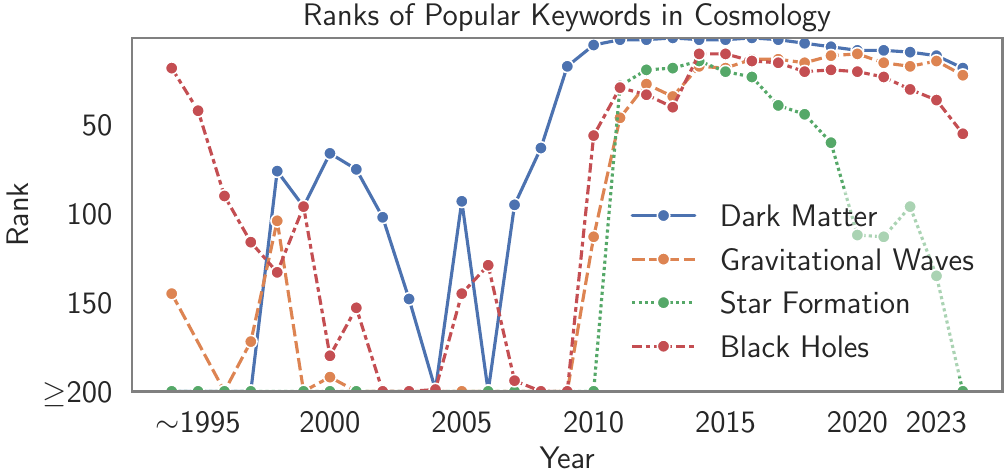}
    \includegraphics[width=0.49\linewidth]{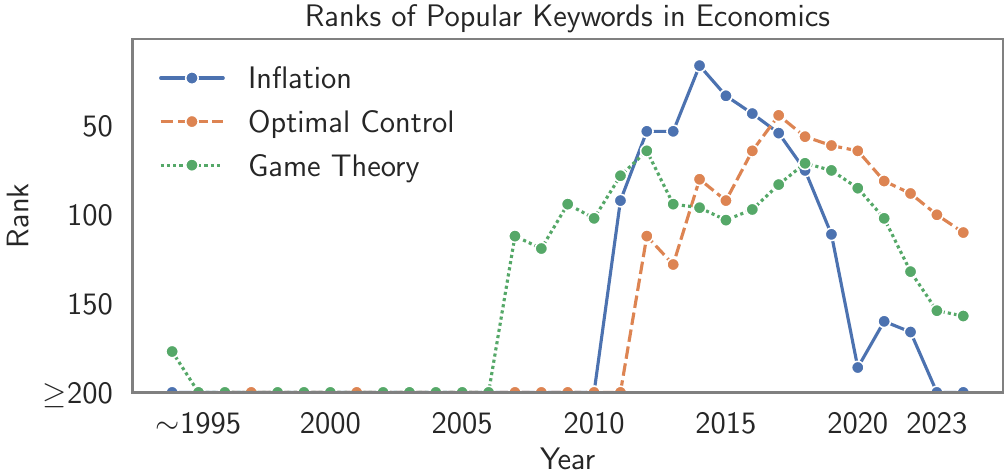}
    \includegraphics[width=0.49\linewidth]{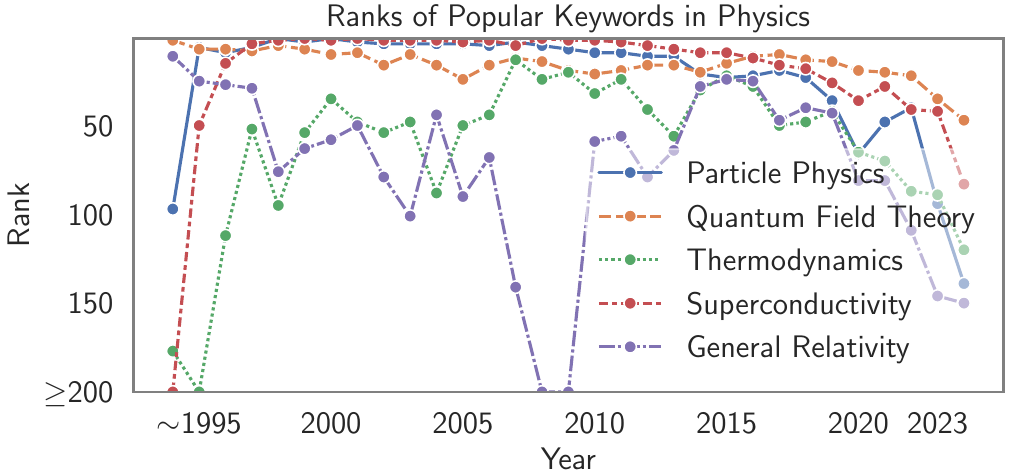}
    \caption{Ranks of mathematics-related (top) and machine-learning-related (bottom) keywords. }
    \label{fig:keyword_ranks_all_others}
\end{figure*}

\begin{table*}
\small
\centering
\begin{tabular}{l|c}
\toprule
\textbf{Year} & \textbf{Closest Keywords to \texttt{Machine Learning}} \\
\midrule
$\sim$ 1995 & Information Retrieval, Classification, Neural Networks, POS Tagging\\
2000 & Ensembles, Logic, Optimization, Machine Translation \\
2005 & Regularization, Support Vector Machines, Data Mining, Reinforcement Learning \\ 
2010 & Image Classification, Kernel Methods, Reinforcement Learning, Transfer Learning \\
2015 & Object Recognition, Question Answering, Image Generation\\
2020 & Ethics, Scalability, Model Explanation, Post-hoc \\
2022 & Data-driven Analysis, Performance Improvement, Medical Imaging, Cross-lingual \\
2024 & General AI, Large Language Models, Conversational Agents, Retrieval-augmented Generation \\
\bottomrule
\end{tabular}
\caption{Closest keywords to \texttt{Machine Learning} in the embedding space, reflecting the shifting focus of academic discourse. }
\label{tab:closest_ml_keywords}
\end{table*} 
\begin{table*}[htbp]
    \centering
    \small
    \begin{tabularx}{1.0\textwidth}{l|X}
    \toprule
        Time Period & Top 5 Most Mentioned Keywords across All Subjects \\
        \midrule
    $\sim$ 1994 & Algebraic Geometry, Quantum Field Theory, Quantum Groups, Quantum Gravity, Lattice QCD \\
    1995 --- 2004 & Superconductivity, Algebraic Geometry, Lattice QCD, Particle Physics, Quantum Mechanics \\
    2005 --- 2014 & Algebraic Geometry, Superconductivity, Quantum Mechanics, Dark Matter, Cosmology \\
    2015 --- 2017 & Deep Learning, Dark Matter, Graph Theory, Cosmology, Optimization \\
    2018 --- 2019 & Deep Learning, Machine Learning, Neural Networks, Dark Matter, Reinforcement Learning \\
    2020 --- 2021 & Deep Learning, Machine Learning, Transformers, Reinforcement Learning, COVID-19 \\
    2022 --- 2023 & Deep Learning, Machine Learning, Transformers, Reinforcement Learning, Quantum Computing \\
    2024 $\sim$ & Machine Learning, Language Models, Large Language Models, Deep Learning, Reinforcement Learning \\
    \bottomrule
    \end{tabularx}
    \caption{Most frequently mentioned keywords in paper titles across different time periods. arXiv shows a noticeable shift in epistemic priority from mathematics and physics to computer science topics.}
    \label{tab:most_mentioned_keywords}
\end{table*}

\begin{table*}[htbp]
    \centering
    \small
    \begin{tabularx}{1.0\textwidth}{l|X}
    \toprule
        Time Period & Top 5 Most Mentioned Keywords in Computer Science\\
        \midrule
    $\sim$ 1995 & Parsing, Machine Translation, Linguistics, Syntax, Semantics \\
    1996 --- 2000 & Parsing, HPSG, ATM Networks, TCP, Linguistics\\
    2001 --- 2005 & Optimization, Logic Programming, Computational Complexity, Quantum Computing \\
    2006 --- 2010 & Information Theory, Graph Theory, Optimization, MIMO, Wireless Networks \\
    2011 --- 2015 & Optimization, Graph Theory, Machine Learning, Complexity, Information Theory \\
    2016 --- 2020 & Deep learning, Machine Learning, Self-supervised Learning, Reinforcement Learning, Optimization \\
    2021 --- 2022 & Deep Learning, Machine Learning, Transformers, Reinforcement Learning, COVID-19 \\
    2023 --- 2024 & Large Language Models, Deep Learning, Machine Learning, Reinforcement Learning, Fairness \\
    \bottomrule
    \end{tabularx}
    \caption{Most frequently mentioned keywords in the field of Computer Science.}
    \label{tab:most_mentioned_cs_keywords}
\end{table*}

\begin{table*}[htbp] 
\centering
\small
\begin{tabularx}{1.0\textwidth}{l|X} 
\toprule Time Period & Top 5 Most Mentioned Keywords in Economics Literature \\ 
\midrule 
2005 --- 2009 & Quantile Regression, Cornish-Fisher Expansion, Instrumental Variables, Panel Data, Econometrics \\
2010 --- 2014 & Machine Learning, Causal Inference, Game Theory, Nonparametric Methods, Forecasting \\
2015 --- 2019 & Machine Learning, Causal Inference, Game Theory, Decision Making, Economics \\
2020 --- 2022 & COVID-19, Machine Learning, Economic Growth, Inequality, Forecasting \\ 
2023 --- 2024 & Artificial Intelligence, Reinforcement Learning, Economics, Mechanism Design, Inequality \\ 
\bottomrule
\end{tabularx} 
\caption{Top 5 most mentioned keywords in economics research over time.} \label{tab:most_mentioned_economics_keywords} \end{table*}

\begin{table*}[!ht]
    \centering
    \begin{tabularx}{1.0\textwidth}{l|X}
    \toprule
        Subject & Category \\
        \midrule
        Computer Science (cs) & cs.AI, cs.AR, cs.CC, cs.CE, cs.CG, cs.CL, cs.CR, cs.CV, cs.CY, cs.DB, cs.DC, cs.DL, cs.DM, cs.DS, cs.ET, cs.FL, cs.GL, cs.GR, cs.GT, cs.HC, cs.IR, cs.IT, cs.LG, cs.LO, cs.MA, cs.ML, cs.MM, cs.MS, cs.NA, cs.NE, cs.NI, cs.OH, cs.OS, cs.PF, cs.PL, cs.RO, cs.SC, cs.SD, cs.SE, cs.SI, cs.SY \\ 
        \midrule
        Economics (econ) & econ.EM, econ.GN, econ.TH \\ 
        \midrule
        {Electrical Engineering}
        & \multirow{2}{*}{eess.AS, eess.IV, eess.SP, eess.SY} \\ 
        and Systems Science (eess) & \\
        \midrule
        Mathematics (math) & math.AC, math.AG, math.AP, math.AT, math.CA, math.CO, math.CT, math.CV, math.DG, math.DS, math.FA, math.GM, math.GN, math.GR, math.GT, math.HO, math.IT, math.KT, math.LO, math.MG, math.MP, math.NA, math.NT, math.OA, math.OC, math.PR, math.QA, math.RA, math.RT, math.SG, math.SP, math.ST \\ 
        \midrule
        Physics (physics) & astro-ph.CO, astro-ph.EP, astro-ph.GA,  astro-ph.HE, astro-ph.IM, astro-ph.SR,  cond-mat.dis-nn, cond-mat.mes-hall,  cond-mat.mtrl-sci, cond-mat.other,  cond-mat.quant-gas, cond-mat.soft,  cond-mat.stat-mech, cond-mat.str-el,  cond-mat.supr-con, gr-qc, hep-ex, hep-lat,  hep-ph, hep-th, math-ph, nlin.AO, nlin.CD,  nlin.CG, nlin.PS, nlin.SI, nucl-ex,  nucl-th, physics.acc-ph, physics.ao-ph,  physics.app-ph, physics.atm-clus,  physics.atom-ph, physics.bio-ph,  physics.chem-ph, physics.class-ph,  physics.comp-ph, physics.data-an,  physics.ed-ph, physics.flu-dyn,  physics.gen-ph, physics.geo-ph,  physics.hist-ph, physics.ins-det,  physics.med-ph, physics.optics,  physics.plasm-ph, physics.pop-ph,  physics.soc-ph, physics.space-ph, quant-ph \\ 
        \midrule
        Quantitative Biology (q-bio) & q-bio.BM, q-bio.CB, q-bio.GN, q-bio.MN, q-bio.NC, q-bio.OT, q-bio.PE, q-bio.QM, q-bio.SC, q-bio.TO \\ 
        \midrule
        Quantitative Finance (q-fin) & q-fin.CP, q-fin.EC, q-fin.GN, q-fin.MF, q-fin.PM, q-fin.PR, q-fin.RM, q-fin.ST, q-fin.TR \\ 
        \midrule
        Statistics (stat) & stat.AP, stat.CO, stat.ME, stat.ML, stat.OT, stat.TH \\ 
        \bottomrule
    \end{tabularx}
    \caption{The arXiv Taxonomy contains 8 major subject areas, including computer science (cs), economics (econ), electrical engineering and systems science (eess), mathematics (math), quantitative biology (q-bio), quantitative finance (q-fin), statistics (stat), and physics.}
    \label{tab:arXivTaxonomy}
\end{table*}




\subsection{Hyperparameters}
For keyword trajectory generation (Section~\ref{sec:micro_level}), we train the Graph Convolutional Networks (GCN) model for 50 epochs using the Adam optimizer with an initial learning rate of 0.01, along with a linear decay learning rate scheduler.  We apply a 1:1 negative sampling ratio to balance positive and negative edges. For t-SNE visualization, we set the perplexity to 30 and run the optimization for a maximum of 1000 iterations.

\subsection{Usage of AI Assistants}

We use GPT-4o to improve the writing of our manuscript.

\section{Data Release Plan}

We plan to release our dataset on Zenodo 
to ensure long-term  access. 
To comply with arXiv's policy prohibiting third-party hosting of e-prints, we provide scripts for downloading the PDF e-prints directly from arXiv. 

\begin{figure*}[htbp]
    \centering
    \includegraphics[width=0.97\linewidth]{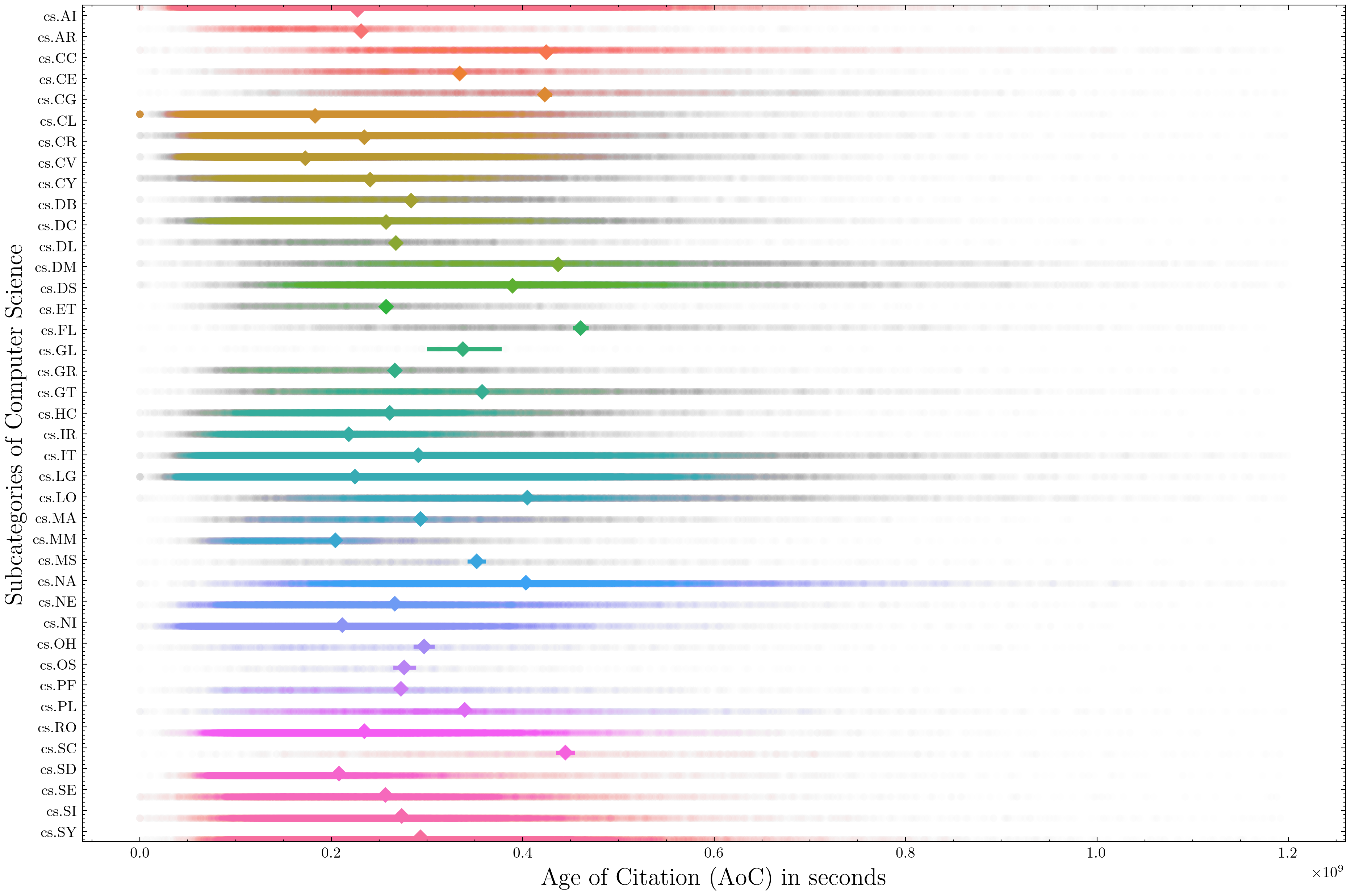}
    \caption{Age of Citation (AoC) of papers under different subfields in CS. The median AoC is marked in diamond shape.
    }
    \label{fig:aoc_cs}
\vspace{-2mm}
\end{figure*}

\section{Motivation for Selecting arXiv}
\label{app:data_selection}
We select \href{https://arxiv.org/}{arXiv} as the data source due to the following reasons:
\begin{itemize}[leftmargin=1em]
\setlength\itemsep{0em}
    \item \emph{Widespread Adoption}: arXiv is the go-to platform for researchers, especially in technical fields such as physics, computer science, and mathematics, where it has become a standard for disseminating preprints.
    \item \emph{Community Trust and Early Impact.} Unlike papers published in conferences or journals, which often involve lengthy peer review processes, arXiv allows researchers to share their findings quickly as preprints. This makes arXiv a critical resource for accessing the latest scientific developments in real-time and gauging when breakthrough technologies appear. arXiv papers often have significant early impact, as many researchers use the platform to gauge emerging trends and cite preprints in their work even before formal publication. 
    \item \emph{Multidisciplinary Coverage.} As shown in the arXiv Taxonomy in Figure~\ref{tab:arXivTaxonomy}, arXiv supports a wide array of fields, making it an ideal platform for cross-domain studies and the exploration of interdisciplinary trends.
    \item \emph{Data Permanence and Integrity.} The permanence of papers on arXiv means that once they are part of the dataset, they remain available indefinitely. This is unlike other platforms such as ResearchGate\footnote{\url{https://www.researchgate.net/}}, which allow authors to remove or significantly modify their works. As arXiv papers cannot be deleted, citation networks based on arXiv papers remain intact and reliable. This is crucial for scientometric analysis that requires accurate and stable citation relationships.
    \item \emph{Versioning} Authors can submit updated versions of their work, reflecting changes, corrections, or new findings. This versioning system provides a clear historical record of how a paper evolves without losing the original submission. In contrast, other platforms may not track versions as clearly or may allow full deletion, which can obscure the evolution of research and make it difficult to trace the development of scientific ideas over time.
\end{itemize}

\begin{figure}[htbp]
\centering
\vspace{-2mm}
\includegraphics[width=0.4\linewidth]{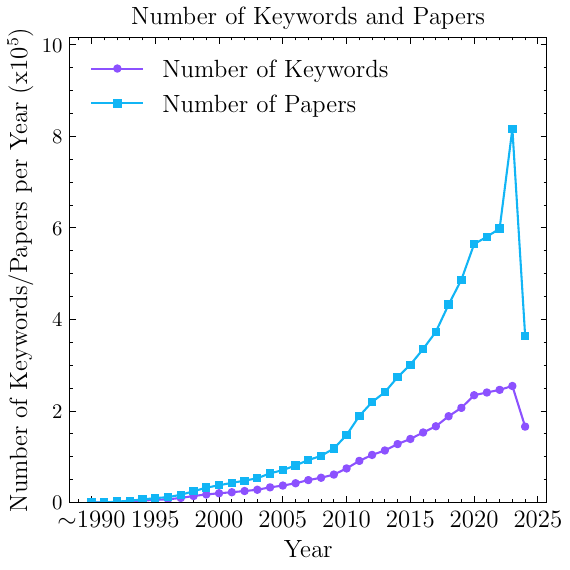}
\caption{Number of papers per year \dataset and their associated keywords. Note that the papers are collected up until June 2024.}
\label{fig:num_papers_and_keywords}
\vspace{-3mm}
\end{figure}

We also acknowledge that the resulting dataset is mainly English-centric. Future works can incorporate non-English academic literature for a more holistic understanding of global academic landscape.


\section{Broader Impact}
The \dataset provides a comprehensive resource for analyzing academic publications over the \emph{past} 30 years, understanding the \emph{present} spatio-temporal knowledge dissemination, and predicting \emph{future} research trends and emerging fields.

\noindent \textbf{Identifying Emerging Fields and Influential Papers.} 
\dataset can be used to identify emerging disciplines, cross-domain collaborations, and effects of global events on shifting academic focus. 
Researchers can also use early citation patterns provided in \dataset to predict influential papers and topics, providing insights into the trajectory of scientific discoveries and intellectual influence.

\noindent \textbf{Fine-grained Citation Analysis.}
By analyzing the text of citing sentences, researchers can perform fine-grained citation analysis to determine the purpose and sentiment of citations. This enables understanding of whether current works are more critical or supportive of new ideas. 

\noindent \textbf{Enhancing Research Discoverability and Knowledge Retrieval.} 
With its structured abstracts, titles, and keywords, \dataset serves as a rich resource for training and benchmarking NLP models in tasks such as text classification, summarization, and keyword extraction. This can significantly enhance research discoverability, support automated literature reviews, and contribute to more efficient knowledge retrieval. Future work can explore technological impacts on academic focus, facilitating the automatic discovery of scientific findings from existing literature~\citep{li2024simulating}.

\noindent \textbf{Studying Spatio-Temporal Knowledge Dissemination.} 
The dataset enables the analysis of spatio-temporal dissemination patterns of scientific knowledge, providing insights into how research ideas originate in specific geographical or subject areas and spread globally over time. 

\noindent \textbf{Inspiring New Research Questions.} According to the theory of questions and question asking, answers to existing questions often give rise to new ones~\citep{ram1991theory}. By examining existing studies within the dataset, researchers can identify research gaps and inspire new questions and directions.

\noindent \textbf{Automatic Survey Generation and Literature Retrieval.} \dataset can be utilized to automatically generate surveys~\citep{wang2024autosurvey}, aiding in the synthesis of knowledge and identification of research gaps. It enables fast literature search and retrieval through keywords~\citep{xiong2024search}, improving access to relevant research and facilitating scholarly communication.

\end{document}